\documentclass[11pt,british]{article}
\usepackage[utf8]{inputenc}
\usepackage{geometry}
\usepackage{color}
\usepackage{babel}
\usepackage{verbatim}
\usepackage{mathtools}
\usepackage{amsmath}
\usepackage{amsthm}
\usepackage{amssymb}
\usepackage{graphicx}
\usepackage{setspace}
\usepackage{microtype}
\usepackage[unicode=true,pdfusetitle,
 bookmarks=true,bookmarksnumbered=false,bookmarksopen=false,
 breaklinks=false,pdfborder={0 0 0},pdfborderstyle={},backref=false,colorlinks=true]
 {hyperref}
\hypersetup{
 colorlinks=true,urlcolor=darkblue,anchorcolor=darkblue,citecolor=darkblue,filecolor=darkblue,linkcolor=darkblue,menucolor=darkblue,pagecolor=darkblue,linktocpage=true}

\makeatletter
\usepackage{tikz}
\usetikzlibrary{backgrounds}
\usepackage[framemethod=tikz]{mdframed}
\AtBeginEnvironment{mdframed}{%
\tikzset{every picture/.style={}}%
}
\mdfsetup{roundcorner=.5ex}
{\begin{mdframed}[backgroundcolor=white]}%
{\end{mdframed}}

\usepackage{slashed}	 	
\usepackage{amsfonts} 		
\usepackage{cite}	
\SetTracking{encoding={*}, shape=sc}{0} 	
\usepackage{color} 		
\definecolor{darkblue}{rgb}{0.0,0.0,0.3} 	
\allowdisplaybreaks		
\date{\today} 		
\numberwithin{equation}{section}	
\mathtoolsset{showonlyrefs}		
\usepackage[small]{titlesec}

\makeatletter
\g@addto@macro\bfseries{\boldmath}
\makeatother

\usepackage{xcolor}
\definecolor{darkblue}{rgb}{0.0,0.0,0.4}

\makeatletter
\g@addto@macro\@floatboxreset\centering
\makeatother

\let\originalleft\left
\let\originalright\right
\renewcommand{\left}{\mathopen{}\mathclose\bgroup\originalleft}
\renewcommand{\right}{\aftergroup\egroup\originalright}

\renewenvironment{thebibliography}[1]{%
\begin{oldthebibliography}{#1}%
\small%
\raggedright%
\setlength{\itemsep}{0pt}%
}%
{%
\end{oldthebibliography}%
}

\makeatother

\begin{document}
\global\long\def\rep#1{\boldsymbol{#1}}%
\global\long\def\repb#1{\overline{\boldsymbol{#1}}}%
\global\long\def\dd{\text{d}}%
\global\long\def\ii{\text{i}}%
\global\long\def\ee{\text{e}}%
\global\long\def\Dorf{L}%
\global\long\def\tDorf{\hat{L}}%
\global\long\def\GL#1{\text{GL}(#1)}%
\global\long\def\Orth#1{\text{O}(#1)}%
\global\long\def\SO#1{\text{SO}(#1)}%
\global\long\def\Spin#1{\text{Spin}(#1)}%
\global\long\def\Symp#1{\text{Sp}(#1)}%
\global\long\def\Uni#1{\text{U}(#1)}%
\global\long\def\SU#1{\text{SU}(#1)}%
\global\long\def\Gx#1{\text{G}_{#1}}%
\global\long\def\Fx#1{\text{F}_{#1}}%
\global\long\def\Ex#1{\text{E}_{#1}}%
\global\long\def\ExR#1{\text{E}_{#1}\times\mathbb{R}^{+}}%
\global\long\def\ex#1{\mathfrak{e}_{#1}}%
\global\long\def\gl#1{\mathfrak{gl}_{#1}}%
\global\long\def\SL#1{\text{SL}(#1)}%
\global\long\def\Stab{\operatorname{Stab}}%
\global\long\def\vol{\operatorname{vol}}%
\global\long\def\tr{\operatorname{tr}}%
\global\long\def\ad{\operatorname{ad}}%
\global\long\def\ext{\Lambda}%
\global\long\def\AdS#1{\text{AdS}_{#1}}%
\global\long\def\op#1{\operatorname{#1}}%
\global\long\def\im{\operatorname{im}}%
\global\long\def\re{\operatorname{re}}%
\global\long\def\eqspace{\mathrel{\phantom{{=}}{}}}%
\global\long\def\bZ{\mathbb{Z}}%
\global\long\def\bC{\mathbb{C}}%
\global\long\def\bP{\mathbb{P}}%
\global\long\def\bR{\mathbb{R}}%
\global\long\def\feyn#1{\slashed{#1}}%
\global\long\def\id{\operatorname{id}}%
\global\long\def\ap{\alpha'}%
\global\long\def\del{\partial}%
\global\long\def\bdel{\bar{\partial}}%


\begin{center}
{\setstretch{1.3}\Large\bf Calabi--Yau metrics, CFTs and random matrices\par} 
\vskip .5cm 
Anthony Ashmore
\vskip .5cm
\textit{\small Enrico Fermi Institute \& Kadanoff Center for Theoretical Physics,\\ University of Chicago, Chicago, IL 60637, USA}  \\[.2cm] \textit{\small Sorbonne Universit\'e, CNRS, LPTHE, F-75005 Paris, France}
\end{center}
\begin{center}\small{Based on a talk given at the Nankai Symposium on Mathematical Dialogues 2021}\end{center}
\begin{center} \small{\textbf{Abstract}} \end{center}
\begin{quote}
\small{Calabi--Yau manifolds have played a key role in both mathematics and physics, and are particularly important for deriving realistic models of particle physics from string theory. Unfortunately, very little is known about the explicit metrics on these spaces, leaving us unable, for example, to compute particle masses or couplings in these models. We review recent progress in this direction on using numerical approximations to compute the spectrum of the Laplacian on these spaces. We give an example of what one can do with this new ``data'', giving a surprising link between Calabi--Yau metrics and random matrix theory.}
\end{quote}

\section{Introduction and summary}

Calabi--Yau manifolds have a rich mathematical history and give a starting point for recovering realistic four-dimensional physics from string theory. Despite much study, there are still no explicit expressions for non-trivial Ricci-flat metrics on these spaces.\footnote{At least in six dimensions and higher -- see \cite{1810.10540, 2006.02435, 2010.12581} for progress on computing K3 metrics.} Viewed from the string worldsheet, these manifolds define interacting two-dimensional superconformal field theories (SCFTs). These theories come in families labelled by the complex structure and Kähler moduli of the underlying Calabi--Yau. In the large-volume limit, these theories are described by non-linear $\sigma$-models whose target space is simply the Calabi--Yau equipped with its Ricci-flat metric.

Recent advances in numerical methods give us access to both the Ricci-flat metric and the spectrum of the Laplacian on these spaces. The spectrum is an source of new non-BPS ``data'' which characterises CFT operators with low scaling dimension. In this talk, we show that the spectrum of the Laplacian, and thus the spectrum of operators in the corresponding CFT, averaged over complex structure moduli displays the hallmarks of chaos in the form of random matrix statistics. In our companion paper~\cite{Afkhami-Jeddi:2021qkf}, we also examine K3 CFTs and the spectra of field theories on genus-three Riemann surfaces, again finding chaos in their spectra.

Further work includes extending our analysis to the non-scalar spectrum, understanding the interplay of the explicit spectrum with the conformal~\cite{Hellerman:2009bu,Keller:2012mr,Friedan:2013cba,Lin:2015wcg,Collier:2016cls,Lin:2016gcl,Kravchuk:2021akc} and geometric \cite{1910.04767,2007.10337,Bonifacio:2021msa,Bonifacio:2021aqf} bootstrap programmes, exploring whether RMT sheds light on interacting CFTs in an averaged sense or if it can be used to understand the ``typical'' properties of Calabi--Yau compactifications~\cite{Ashok:2003gk,Douglas:2004zu,Douglas:2003um,Denef:2004ze,Denef:2004cf,Distler:2005hi,Podolsky:2008du}

A point to emphasise is that one could imagine taking random metrics on these manifolds and finding similar random matrix statistics. Our surprising result is that one sees this chaotic behaviour even within the restricted class of Ricci-flat metrics on a Calabi--Yau hypersurface. A related observation is that one sees similar behaviour for constant negative curvature metrics on genus-three Riemann surfaces. In this context, there are mathematical theorems relating the behaviour of geodesics to chaotic particle motion on these manifolds~\cite{Hedlund,Hopf1,Hopf2,Asnov}. This suggests that something similar should be true for Ricci-flat metrics, though this still appears to be an open question.

\section{Chaos in 2d CFTs from Calabi--Yau \texorpdfstring{$\sigma$}{sigma}-models?}

Even in two dimensions, very little is known about the spectrum of interacting CFTs. %
{} For CFTs at generic points in moduli space, one can study quantities protected by supersymmetry~\cite{Witten:1982df,Eguchi:1987wf,Eguchi:1988vra,Cecotti:1992qh,Kawai:1993jk,Dijkgraaf:1996it,Benini:2013nda,Lin:2015dsa} or use modular invariance and the conformal bootstrap to constrain the spectrum in some way~\cite{Hellerman:2009bu,Keller:2012mr,Friedan:2013cba,Lin:2015wcg,Collier:2016cls,Lin:2016gcl}. In general, however, there is no way to compute the spectrum of an interacting CFT.

There are large families of SCFTs defined by Calabi--Yau manifolds. For example, a $(2,2)$ $\sigma$-model with a Calabi--Yau threefold target space $X$ is known to flow to a $(2,2)$ SCFT~\cite{Hull:1985at,AlvarezGaume:1985xfa,Nemeschansky:1986yx}, with the complex structure and Kähler moduli of the Calabi--Yau metric determining certain exactly marginal couplings in the field theory~\cite{Friedan:1980jf,AlvarezGaume:1981hm,AlvarezGaume:1980dk,AlvarezGaume:1981hn,AlvarezGaume:1985ww,Gross:1986iv}.

The relation of the geometry of the Calabi--Yau to the spectrum of operators in the CFT was given first by Witten~\cite{Witten:1982df}. At low energies, the $\sigma$-model reduces to a quantum mechanics for a point particle moving on the target space, with the Hamiltonian given by the de Rham Laplacian, $\Delta$, for the Ricci-flat metric. Thanks to this, one can study the low-lying operators of the CFT via the geometry of the Calabi--Yau. Primary operators $\{\mathcal{O}_{i}\}$ in the CFT are eigenstates of the Hamiltonian with fixed scaling dimension $D_{i}$:
\begin{equation}
H|\mathcal{O}_{i}\rangle=D_{i}|\mathcal{O}_{i}\rangle,\qquad D_{i}\geq0,
\end{equation}
In the large-volume limit, these operators correspond to the $(p,q)$-eigenforms of $\Delta$, with scaling dimensions
\begin{equation}
D=\lambda+\frac{p+q}{2},
\end{equation}
where $\lambda$ is the eigenvalue of the corresponding eigenform. Since $\lambda\sim\op{Vol}(X)^{-1/\dim_{\bC}X}$, at large volume the light operators come from \emph{scalar} eigenmodes of $\Delta$.

In the spirit of ``experimental'' theoretical physics, we ask the following question: 
\begin{quote}
Given an ensemble of CFTs, do the scaling dimensions $\{D_{i}\}$ of primary operators display any interesting statistics? In particular, do they display signs of chaos?
\end{quote}
This is obviously a difficult question to answer: we need families of generic, interacting CFTs, and we have to be able to compute the spectrum of these theories. The requirement of being generic and interacting immediately rules out looking at solvable or rational CFTs, and needing families of such theories means we cannot look at special points in moduli space or isolated CFTs, such as the Ising model.

Until recently, it was not possible to answer such a question, other than in free field theories~\cite{Afkhami-Jeddi:2020ezh,Maloney:2020nni,2103.15826}. The idea we pursue in this talk is to construct an ensemble of CFTs and compute their spectra numerically via the connection to Calabi--Yau geometries. We then analyse the spectra statistically and find that they display random matrix statistics, indicative of chaos in the underlying field theory.

\section{Numerical Calabi--Yau metrics and their spectra}

Calabi--Yau (CY) manifolds are Kähler manifolds which admit Ricci-flat metrics. Thanks to Yau's proof~\cite{Yau:420951} of the Calabi conjecture~\cite{Calabi57}, we know that such metrics exist when $c_{1}(X)=0$. This proof is in no way constructive, however, and so the metrics must be determined by other means.

There are now many methods to compute numerical approximations to Ricci-flat metrics on CY manifolds~\cite{Headrick:2005ch,Tian,math/0512625,Douglas:2006rr,Braun:2007sn,Headrick:2009jz,Ashmore:2019wzb,Anderson:2020hux,Douglas:2020hpv,Jejjala:2020wcc,Douglas:2021zdn,Larfors:2021pbb,Ashmore:2021ohf}, which have been discussed many times in the literature~\cite{math/0512625,hep-th/0506129,hep-th/0612075,0712.3563,0908.2635,1910.08605,Cui:2019uhy,Ashmore:2020ujw,2012.04656,2012.04797,2012.15821,Ashmore:2021rlc}. In this work, we use an algorithm due to Donaldson~\cite{math/0512625}, based on the \emph{algebraic metrics} ansatz of Tian~\cite{Tian}. One iteratively solves for a \emph{balanced} metric on $X$, which is known to approach the Ricci-flat metric in a certain limit. In practice, this involves converting integrals over $X$ to discrete sums -- one uses Monte Carlo methods to implement this using a sampling method discussed in \cite{hep-th/0612075,0712.3563}.

\subsection{The Laplacian on a Calabi--Yau}

In the large-volume limit, low-lying operators in the CFT are determined by scalar eigenmodes on the CY. These eigenmodes satisfy
\begin{equation}
\Delta\phi_{n}=\lambda_{n}\phi_{n},\qquad\Delta=\delta\dd.
\end{equation}
Given the data of an approximate CY metric, we can then calculate the spectrum $\{\lambda_{n}\}$ following \cite{0804.4555,0805.3689,Ashmore:2020ujw,2103.07472}. We do this by expanding the eigenmodes in a truncated, finite basis of functions $\{\alpha_{A}\}$, so that $\phi=\phi_{A}\alpha_{A}$, thus giving a finite-dimensional generalised eigenvalue problem,
\begin{equation}
\Delta_{AB}\phi_{B}=\lambda\,O_{AB}\phi_{B},
\end{equation}
where $\Delta_{AB}=\langle\alpha_{A},\Delta\alpha_{B}\rangle$ and $O_{AB}=\langle\alpha_{A},\alpha_{B}\rangle$ both depend on the approximate CY metric.

\section{Calabi--Yau CFTs and random matrix theory}

A suitable ensemble of CFTs is provided by constructing a family of CY manifolds with varying complex structure moduli. For example, a generic quintic threefold $X$ is defined by a quintic equation in $\bP^{4}$ with coordinates $[z_{0}\!:\!\dots\!:\!z_{4}]$:
\begin{equation}
X\equiv\sum_{m,n,p,q,r}c_{mnpqr}z_{m}z_{n}z_{p}z_{q}z_{r}=0.\label{eq:quintics}
\end{equation}
This describes a 101-dimensional family of CYs. We sample the $c_{mnpqr}$ randomly from the unit disk in the complex plane with a flat measure:
\begin{equation}
c_{mnpqr}\in\mathbb{C},\qquad|c_{mnpqr}|<1.
\end{equation}
For each choice of complex structure moduli, we compute an approximate Ricci-flat metric and the spectrum of the Laplacian numerically. This gives us an ensemble of spectra which can be interpreted as an ensemble of scaling dimensions of operators in the large-volume CFT.

Since we are looking for signs of chaos in the spectrum of CFTs, it is useful to have some diagnostics of chaotic behaviour. In our case, we take random matrix statistics to be indicative of chaos in the spectrum~\cite{mehta2004random,Br_zin_1997,RevModPhys.53.385,Guhr:1997ve,B1,B2}.

\subsection{Random matrix theory and spectral statistics}

Random matrix theory (RMT) is the study of the distribution of eigenvalues of matrices whose entries are independent and identically distributed random variables. RMT governs many physical systems, such as nuclear physics~\cite{10.2307/1970079,PhysRev.120.1698,PhysRev.123.1293,Haq19821086}, billiards~\cite{B1,Gutzwiller_1990,BALAZS1986109}, quantum many-body systems such as the SYK model~\cite{Sachdev:1992fk,Kitaev1,Kitaev2,Kitaev3}, and toy models of black-hole physics and quantum gravity~\cite{Saad:2018bqo,Garcia-Garcia:2016mno,You_2017,Cotler:2016fpe,Gharibyan:2018jrp}. The latter suggests, via holography, that generic conformal field theories may also display signs of chaos.

We will compare the statistics of scaling dimensions of primary operators with the eigenvalue statistics of a random matrix theory, specifically the Gaussian orthogonal ensemble (GOE). This is the ensemble of eigenvalues of $N\times N$ real, symmetric matrices, where the entries of the matrices are independently drawn from a Gaussian distribution. We will focus on universal features of random matrix theory by \emph{unfolding}, letting us focus on the statistics of fluctuations in the spectrum.

We measure RMT behaviour via three diagnostics: the nearest-neighbour level spacing $p_{1}(s)$, the number variance $\Sigma^{2}(L)$, and the spectral form factor $S(t,\beta)$. Together, these provide measures of both long- and short-range correlations in the spectrum~\cite{mehta2004random,Guhr:1997ve}.

\begin{figure}
\begin{centering}
\includegraphics{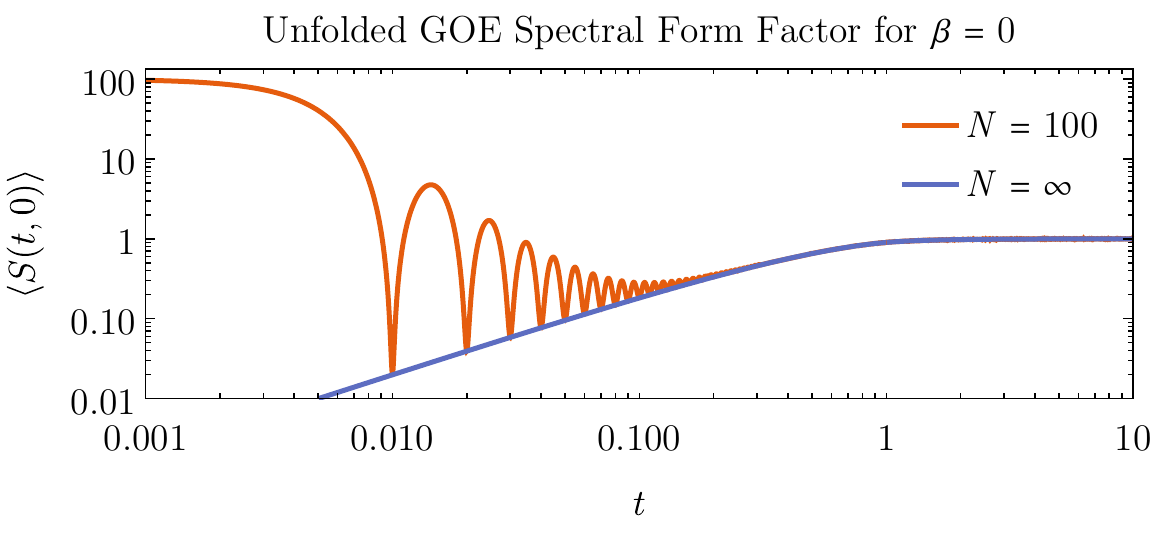}
\par\end{centering}
\caption{Unfolded GOE spectral form factors for $\beta=0$. The orange curve is computed from 100,000, $100\times100$ random symmetric matrices. The blue curve shows the SFF for $N\to\infty$ and $\beta\to0$ -- only the ramp and plateau are present.}
\label{fig:goe_sff}

\end{figure}

The nearest-neighbour level spacing is the probability of two consecutive eigenvalues being separated by a distance $s$. For the GOE in the large-$N$ limit, it is
\begin{equation}
p_{1}(s)=\frac{\pi}{2}s\,\ee^{-\pi s^{2}/4}.
\end{equation}
The maximum of this distribution is away from $s=0$, indicative of \emph{eigenvalue repulsion}. 
The number variance is the fluctuation of the number of eigenvalues averaged over an interval of size $L$. For the GOE, it scales as
\begin{equation}
\Sigma^{2}(L)\sim\log L.
\end{equation}
The logarithmic growth is an indication of \emph{spectral rigidity}. 
The spectral form factor (SFF) (up to an overall normalisation) is the analytically continued thermal partition~\cite{Haake10}
\begin{equation}
S(t,\beta)\sim\biggl|\sum_{n}\ee^{-(\beta+2\pi\ii t)\lambda_{n}}\biggr|^{2}.
\end{equation}
The spectral form factor displays three characteristic features: a dip with ringing for finite $N$, a ramp, and a plateau. We show these features for the GOE in Figure \ref{fig:goe_sff}.

\subsection{Results}

We generated 1,000 different samples of quintic threefolds of the form (\ref{eq:quintics}), with complex structure moduli given by random choices of $c_{mnpqr}$. For each sample, we keep the lowest-lying 100 eigenvalues of the Laplacian. We then repeat this for 1,000 different choices of the $c_{mnpqr}$, and analyse the resulting distribution of operator scaling dimensions.

\begin{figure}
\begin{centering}
\includegraphics{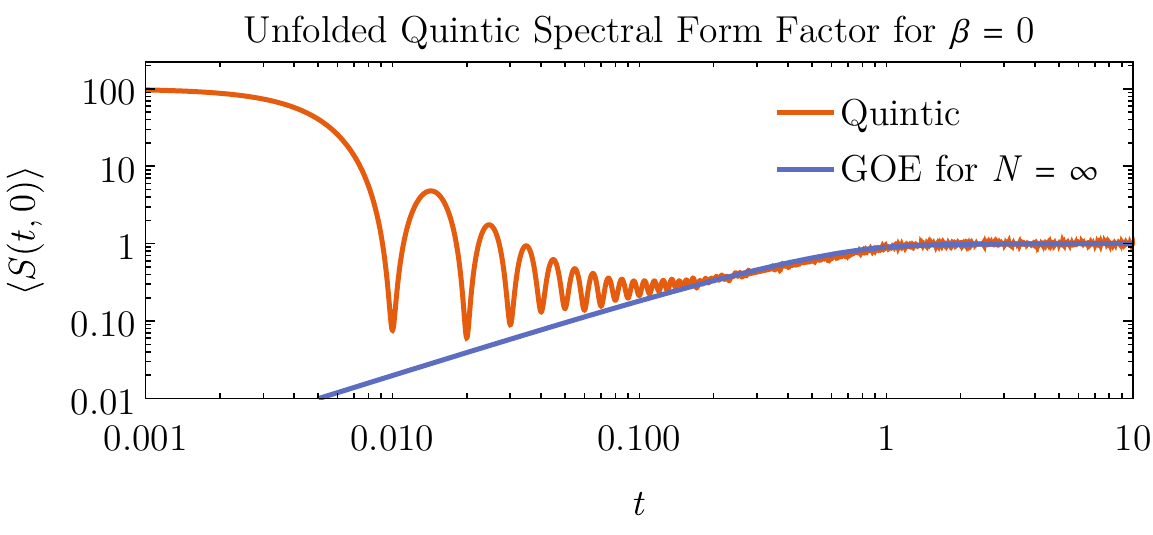}\\
\includegraphics{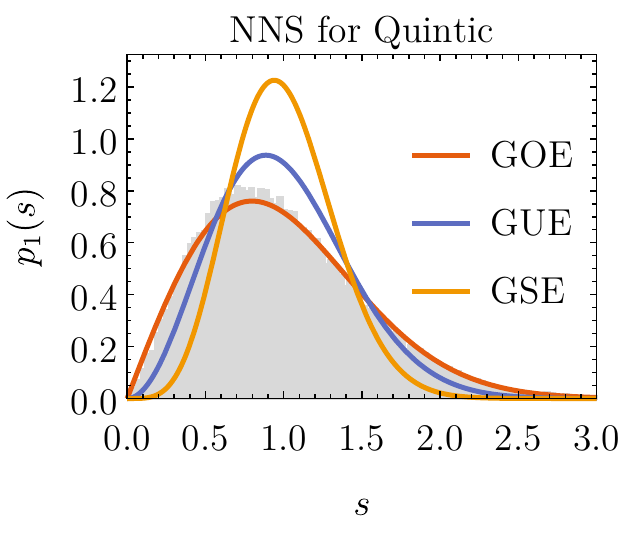}\quad{}\includegraphics{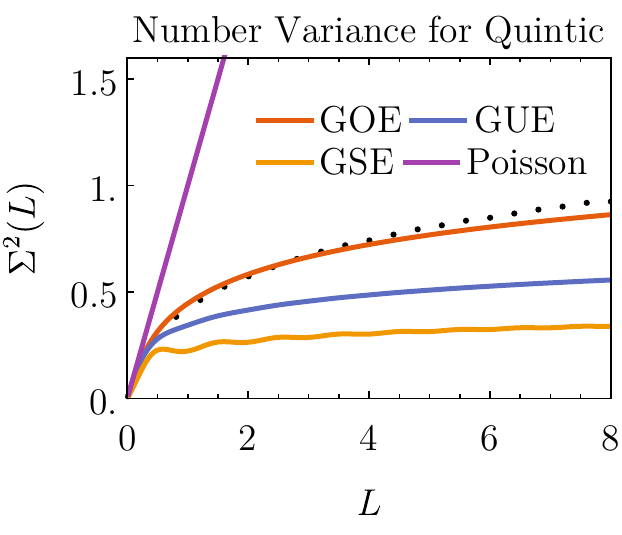}
\par\end{centering}
\caption{The first row shows the unfolded spectral form factor for $\beta=0$ for quintic Calabi--Yau threefolds. The orange curve shows an ensemble average consisting of 1,000 samples with varying complex structure, with each sample containing the 100 lowest-lying eigenvalues. The dip, ramp and plateau are all present. The second row shows the nearest-neighbour level spacings (NNS) and the number variance, together with fits to GOE. Other matrix ensembles are shown for comparison.}
\label{fig:qf}
\end{figure}

We compare the distribution of scaling dimensions with the distribution of eigenvalues from a random matrix ensemble in Figure \ref{fig:qf}. The SFF shows the characteristic dip with ringing, ramp and plateau. The ramp shows a smooth transition to the plateau region, suggesting GOE statistics. The nearest-neighbour level spacing and number variance also match those of the GOE.

\subsection*{Acknowledgements}
We thank Nima Afkhami-Jeddi and Clay C\'ordova for collaboration on the project which this article is based upon. AA is supported by the European Union's Horizon 2020 research and innovation program under the Marie Sk\l{}odowska-Curie grant agreement No.\,838776. This work was completed in part with resources provided by the University of Chicago Research Computing Center.

\bibliographystyle{utphys}
\bibliography{citations_inspire,biblio}

\end{document}